\documentclass[journal,draftclsnofoot, onecolumn]{IEEEtran}

\usepackage{amsmath, graphics,amssymb,epsfig,subfigure,epstopdf}
\usepackage{multirow, cite}
\usepackage{xcolor,colortbl, soul}

\usepackage{tabularx,booktabs}
\newcommand{\mc}[2]{\multicolumn{#1}{c}{#2}}
\definecolor{Gray}{gray}{0.85}
\definecolor{LightCyan}{rgb}{0.88,1,1}
\definecolor{LightBlue}{rgb}{0.75,0.936,1.00}
\sethlcolor{LightBlue}
\newcolumntype{a}{>{\columncolor{Gray}}c}
\newcolumntype{b}{>{\columncolor{white}}c}

\begin{document}

\title{\huge   Non-Orthogonal Multiple Access in Multi-Cell Networks: Theory,
Performance, and Practical Challenges}
\author{Wonjae Shin, Mojtaba Vaezi, Byungju Lee, David J. Love, Jungwoo Lee, and H. Vincent Poor\vspace{-3mm}
\thanks{W. Shin and J. Lee are with the Department of Electrical and Computer Engineering, Seoul National University, Seoul, Korea (e-mail:
\{wonjae.shin, junglee\}@snu.ac.kr). M. Vaezi and H. V. Poor are with the Department of Electrical Engineering, Princeton University, Princeton, NJ, USA (e-mail: \{mvaezi, poor\}@princeton.edu). B. Lee (corresponding author) and D. J. Love are with School of Electrical and Computer
Engineering, Purdue University, West Lafayette, IN, USA (e-mail:
\{byungjulee, djlove\}@purdue.edu). \textit{(Wonjae Shin and Mojtaba Vaezi contributed equally to this work.)}}}
\markboth{IEEE Communications Magazine, Draft}{SHIN, et al.: Coordinated Beamforming for Multi-Cell MIMO-NOMA}
\maketitle
\begin{abstract}
Non-orthogonal multiple access (NOMA) is a potential enabler for the development of 5G and beyond wireless networks.
By allowing multiple users to share the same time and frequency, NOMA can scale up the number of served users, increase the spectral efficiency,  and improve user-fairness compared to existing orthogonal multiple access (OMA) techniques. While single-cell NOMA has drawn significant attention recently, much less attention has been given to multi-cell NOMA. This article discusses the opportunities and challenges of NOMA in  a multi-cell environment.
As the density of base stations and devices  increases, inter-cell interference becomes a major obstacle in multi-cell networks. As such, identifying techniques that combine interference management approaches with NOMA is of great significance. After  discussing the theory behind NOMA, this paper provides an overview of the current literature and discusses key  implementation and  research challenges, with an emphasis on multi-cell NOMA.
\end{abstract}

\section{What Drives NOMA?}

The next generation of wireless networks will require a paradigm shift in order to support massive numbers of devices with diverse data rate and latency requirements. Particularly, the increasing demand for Internet of Things (IoT) devices poses challenging requirements on 5G wireless systems.
Two key features of 5G are expected to be a latency of 1ms, compared to 10 ms in the 3rd Generation Partnership Project (3GPP) Long-Term Evolution  (LTE), and support for 10 Gbps throughput.

To fulfill these requirements, numerous potential technologies
have been introduced over the last few years.
Among them is non-orthogonal multiple access (NOMA) \cite{saito2013non}, a technique to serve multiple users via a single wireless resource. {NOMA can be realized in the power, code, or other domains} \cite{dai2015non,islam2016power}.
Code domain NOMA uses  user-specific spreading sequences for sharing the entire resource, whereas
power domain NOMA exploits the channel gain differences between
the users for multiplexing via power allocation.
Power domain NOMA can improve wireless communication in the following benefits:

\begin{itemize}
	\item \textit{Massive Connectivity:}
	There appears to be a reasonable consensus that NOMA is essential for massive connectivity. This is because the number of served users in all orthogonal multiple access (OMA) techniques is inherently  limited by  the number of resource blocks. In contrast, NOMA theoretically can serve an \textit{arbitrary} number of users in each resource block by superimposing all users' signals. In this sense, NOMA can be tailored to typical IoT applications where a large number of devices sporadically try to transmit small packets.
	
	\item \textit{Low Latency}:
	Latency requirements for 5G application are rather diverse.
	Unfortunately, OMA cannot guarantee such broad delay requirements
	because no matter how many bits a device wants to transmit the device must wait until an unoccupied resource block is available. On the contrary, NOMA supports flexible scheduling since it can accommodate a \textit{variable} number of devices {depending on the application} that is being used and the perceived quality of service (QoS) of the device.

	\item \textit{High Spectral Efficiency:} NOMA also surpasses OMA in terms of spectral efficiency and user-fairness. As will be seen in Section~\ref{sec:theory}, NOMA is the theoretically \textit{optimal} way of using spectrum for both uplink and downlink communications. This is because every NOMA user can enjoy the whole bandwidth, whereas OMA users are limited to a smaller fraction of spectrum which is inversely proportional to the number of users. In addition, NOMA can also be combined with  other emerging  technologies, such as massive multiple-input multiple-output (MIMO) and mmWave technologies, to further support higher throughput.
\end{itemize}

In view of the above benefits, NOMA has drawn much attention from both academia and industry. However, much of the work in this context is limited to single-cell analysis, where there is no co-channel interference caused by an adjacent base station (BS). To verify the benefits of NOMA in a more realistic setting, it is necessary to consider a multi-cell network. Specifically, as wireless networks get denser and denser, inter-cell interference (ICI) becomes a major obstacle to achieving the benefits of NOMA. In this regard, we consider NOMA in a multi-cell environment for this article.
We first discuss the theory behind NOMA and an overview of the literature of NOMA. We then explain the main implementation issues and research challenges, with particular interest on multi-cell NOMA. Finally, the system-level performance evaluation of multi-cell NOMA solutions will be provided before concluding the article.

\begin{figure*}[t]
	\centering
	\hspace{-.77cm}
	\subfigure[MAC (uplink)  ]{
		\includegraphics[scale=.68]{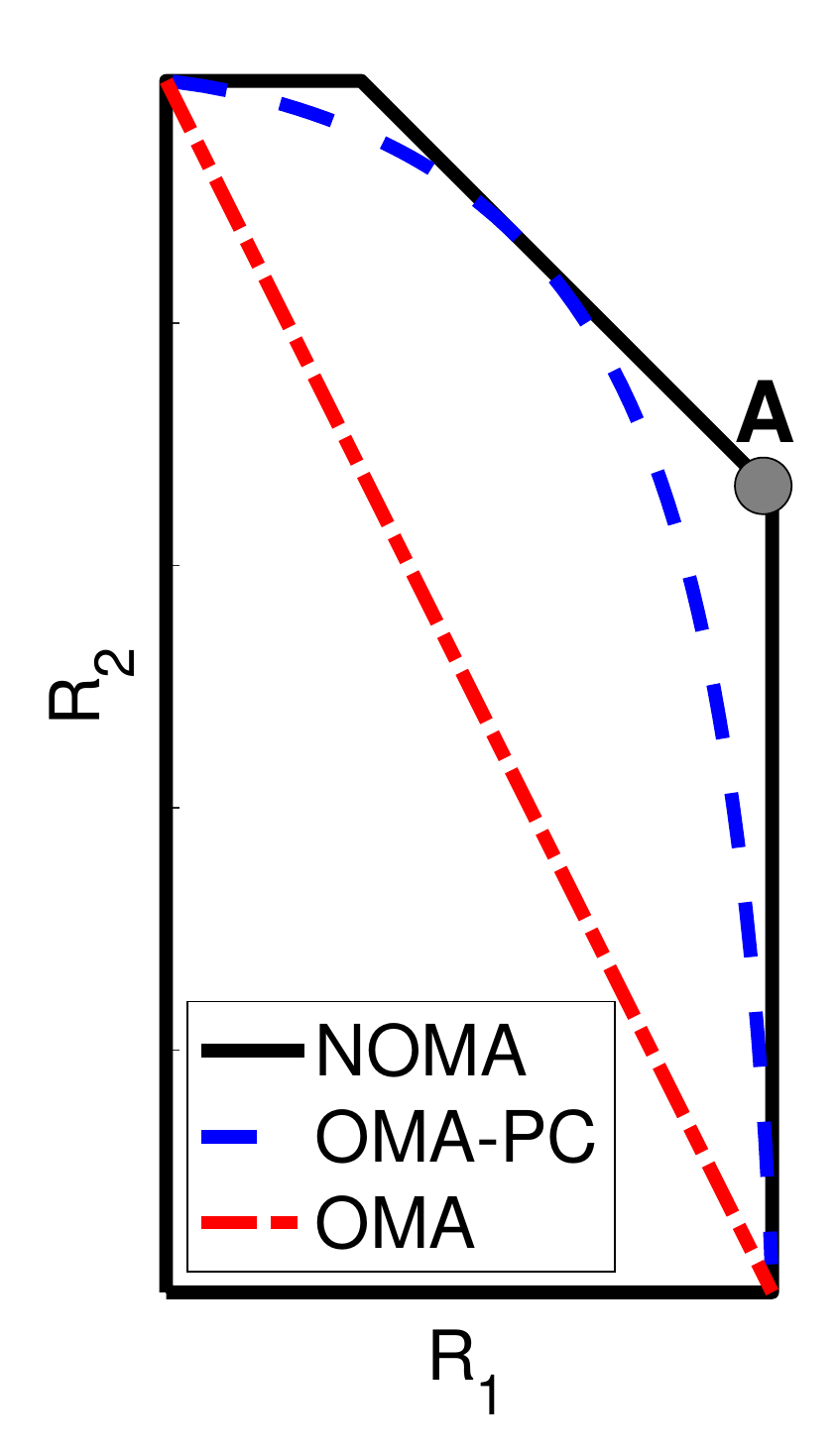}
		\label{fig:MAC}
	} \hspace{-.77cm}
	\subfigure[BC (downlink)]{
		\includegraphics[scale=.68]{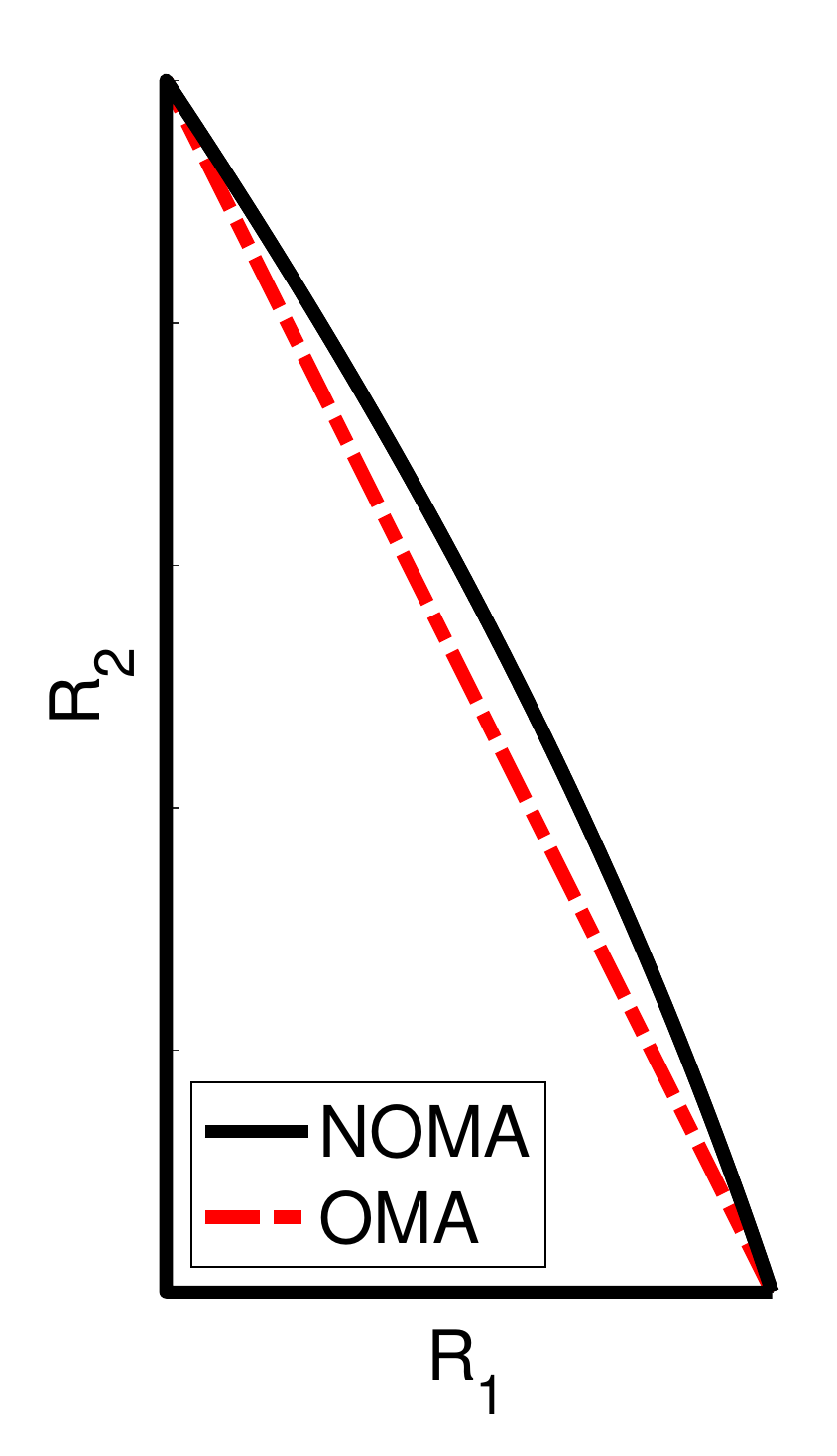}
		\label{fig:BC}
	}\hspace{-.77cm}
	\subfigure[IC (multi-cell)]{
		\includegraphics[scale=.68]{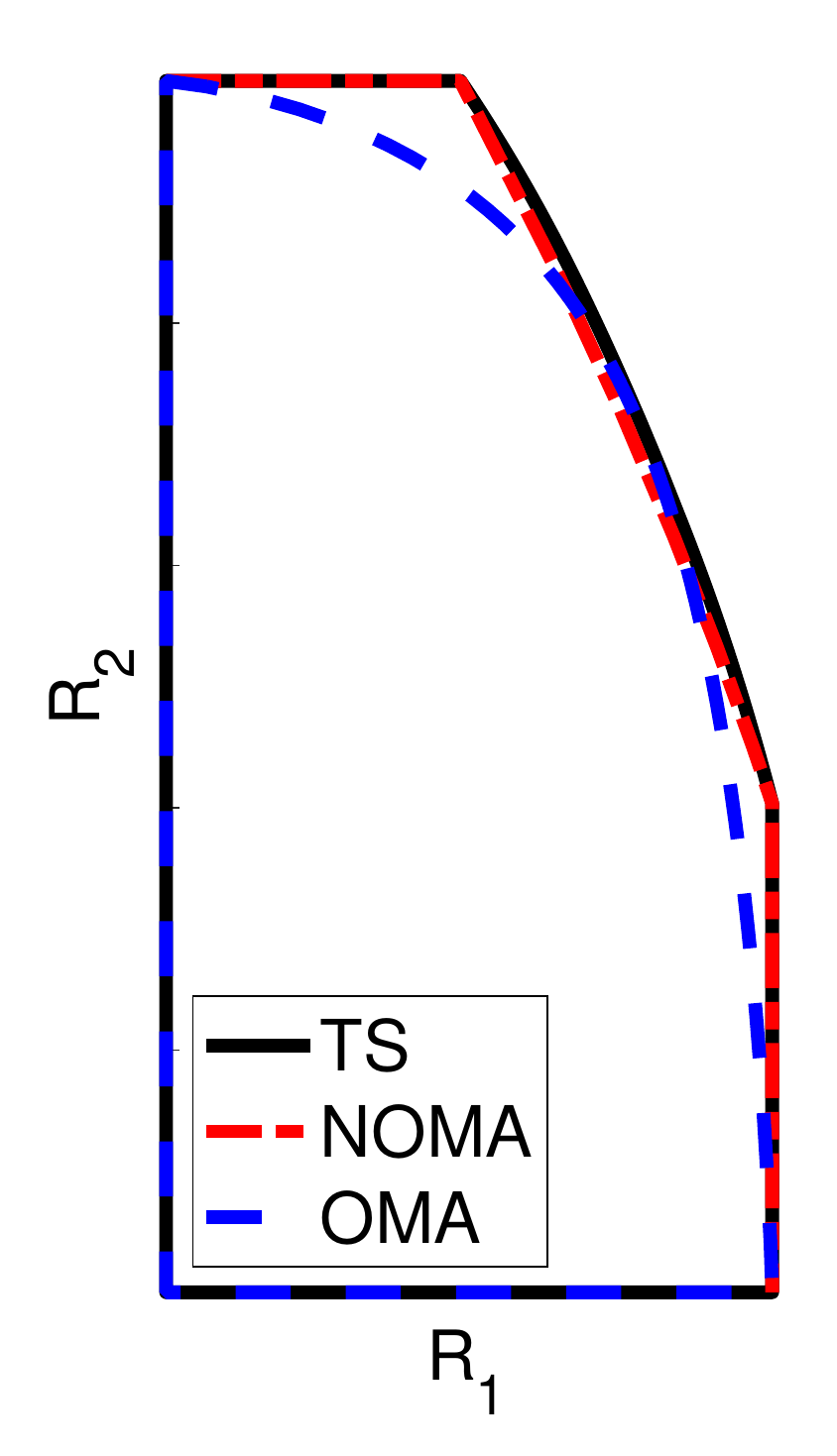}
		\label{fig:IC}
	}
	\caption{
		Best achievable regions by OMA and NOMA in the multiple access channel (MAC), broadcast channel (BC), and interference channel (IC).
	}
	\label{fig:NOMA}
\end{figure*}

\section{Theory Behind NOMA}
\label{sec:theory}

Analysis of cellular communication can generally be classified as either \textit{downlink} or \textit{uplink}. In the downlink channel, the BS simultaneously transmits signals to multiple users, whereas  in the uplink channel multiple users  transmit data to the same BS.

From an information-theoretic perspective, the downlink and uplink are modeled by the \textit{broadcast channel} (BC) and \textit{multiple access channel} (MAC), respectively. The basic premise behind single-cell NOMA in the power domain is to reap the benefits promised by the theory of multi-user channels \cite{tse2005fundamentals}.
As such, we review what information theory promises for these channels,
both in the single-cell and multi-cell settings. In particular, we seek to answer the following two questions in this section: 1) what are the highest achievable throughputs for these multi-user channels? and 2) how can a system achieve such rates?

\subsection{Single-Cell NOMA}
The capacity regions of the two-user MAC and BC are achieved via NOMA, where both users' signals are  transmitted at the same time and in the same frequency band \cite{ElGamal2011network}. The curves labeled by NOMA in Fig.~\ref{fig:MAC} and Fig.~\ref{fig:BC} represent the MAC and BC capacity regions, respectively.
Except for a few points, OMA is strictly suboptimal as can be seen from the figures.
To gain more insight, we describe how the above regions are obtained. For OMA we consider a time division multiple access (TDMA) technique where $\alpha$ fraction of time ($0 \le \alpha\le 1$) is dedicated to user~1 and $\bar \alpha \triangleq  1-\alpha$ fraction of time is dedicated to user~2.
In this paper, $\mathcal{C}(x) \triangleq \frac{1}{2} \log_2(1+x)$ and $\gamma_i = |h_i|^2P$ is the received signal-to-noise ratio (SNR) for user $i$, where $h_i$ is the channel gain, $P$ is the transmitter power, and the noise power is normalized to unity.

\subsubsection{Uplink (MAC)}
Using OMA, each user sees a single-user channel in its dedicated fraction of time and, thus, $ R_1 =\alpha \mathcal{C}(\gamma_1)$ and $R_2 = \bar \alpha \mathcal{C}(\gamma_2)$ are achievable.
If power control is applied, these rates
can be boosted to $ R_1 =\alpha \mathcal{C}(\frac{\gamma_1}{\alpha })$ and
$R_2 = \bar \alpha \mathcal{C}(\frac{\gamma_2}{\bar \alpha })$.
 In  the case of NOMA, both users concurrently transmit, and their signals
 interfere with each other at the BS.
 The BS can use \textit{successive interference cancellation} (SIC) to achieve any point in the NOMA region, which is the capacity region of this channel  \cite{tse2005fundamentals}.
 Particularly, to achieve point $\mathrm{A}$ the BS first decodes user~2's signal treating the other signal as noise. This results in $R_2 = \mathcal{C}(\frac{\gamma_2}{\gamma_1 + 1 })$. The BS then removes user~2's  signal and decodes
user~1's signal free of interference; i.e.,  $ R_1 =  \mathcal{C}(\gamma_1)$.
From Fig.~\ref{fig:MAC} it is seen that the gap between the NOMA and OMA regions becomes larger if power control is not used in OMA.

\subsubsection{Downlink (BC)}
In the downlink, OMA can only achieve $ R_1 = \alpha \mathcal{C}(\gamma_1)$ and $R_2 =\bar \alpha \mathcal{C}(\gamma_2)$.
However, making  use of a NOMA scheme can strictly increase this rate region as shown in Fig.~\ref{fig:BC}.
In particular, the capacity region of this channel is known and can be achieved using
 \textit{superposition coding} at the BS. For decoding, the user with the stronger channel uses SIC to decode its signal free of interference,
i.e.,  $ R_1  = \mathcal{C}(\beta \gamma_1)$, while the other user is
capable of decoding at a rate of $R_2 = \mathcal{C}(\frac{\bar \beta \gamma_2}{\beta \gamma_2 + 1})$,
where $\beta$ is the fraction of the BS power allocated to user~1's data and $\bar \beta = 1- \beta $.
By varying $\beta$ from $0$ to $1$,  any rate pair ($R_1,R_2$) on the boundary of
capacity region of the BC (NOMA region) can be achieved.

The fact that the capacity region of downlink NOMA is known enables us to find the optimum power allocation corresponding to any point ($R_1,  R_2$) on the boundary of the capacity region.
In fact, all we need to know to achieve such a rate pair is to find what fraction of the BS power should be allocated to each user. Corresponding to each ($R_1,  R_2$) there is a $0\le \beta \le 1$
such that $\beta P$ and $\bar \beta P$ are the optimal powers for user~1 and
user~2, respectively, where $P$ is the BS power. Conversely, every $\beta $ generates a point on the boundary of the capacity region.

The above  argument implies that NOMA can improve \textit{user-fairness} smoothly and in an optimal way by flexible power allocation.
Suppose that a  user has a poor channel condition. 
To boost this user's rate and improve user-fairness
the BS can simply  increase the fraction of power allocated to this user.
We can look at this problem from yet another perspective.
To increase the rate of such a user, we can maximize the weighted sum-rate $\mu R_1 + R_2$
where a high weight ($\mu $) is given to such a user.
This is because, to maximize $\mu R_1 + R_2$ for any $\mu\ge0 $
there exists an optimal power allocation strategy, determined by $\beta$.
Seeing that $\mu > 1$ ($\mu < 1$) corresponds to
the case where user~1 has higher (lower) weight than user~2,
to improve the user-fairness we can assign an appropriate weight to the important user  and find
the corresponding $\beta$.

\subsubsection{$K$-User Uplink/Downlink}
In the above, we described coding strategies for the two-user uplink/downlink channels.
Interestingly, very similar coding schemes  are still
capacity-achieving  for the $K$-user MAC and BC, {i.e., superposition coding with SIC gives the largest region for the $K$-user BC.
Similarly, to achieve the capacity region of the $K$-user MAC, the users transmit their signals concurrently and the BS applies SIC, as described in} \cite[Section 6.1.4]{tse2005fundamentals}.
These schemes  are based on NOMA as they allow multiple users to transmit at the same time and frequency.
Additionally, OMA is strictly suboptimal  \cite{tse2005fundamentals}.

\subsection{Multi-Cell NOMA}
In a multi-cell setting, these problems are more involved and simple
channel models are insufficient. Unfortunately,  capacity-achieving schemes are
unknown. However, the achievable rate regions for the interference
channel  indicate the superiority of NOMA to OMA, as shown in Fig.~\ref{fig:IC}.

\subsubsection{Interference Channel (IC)}
The capacity region of the two-user IC is not known in general; however, it is  known that OMA is strictly sub-optimal.
The Han-Kobayashi (HK) scheme \cite{ElGamal2011network} is the best known achievable scheme for the IC.
In its basic form, the HK scheme employs rate-splitting and superposition coding at each transmitter.
Since it uses superposition coding, the basic HK implies a NOMA.
In general, the HK scheme applies time-sharing to improve the basic HK region and can be seen as a combination of NOMA and OMA \cite{vaezi2016simplified}.
The HK scheme that combines
NOMA and OMA gives the largest rate region \cite{vaezi2016simplified}, as shown in Fig.~\ref{fig:IC}.
In this figure, OMA refers to  TDMA whereas NOMA refers to the basic HK scheme
in which  time-sharing is not applied.
The third curve, labeled TS,
is  based on the HK scheme with time-sharing (TS) in which two time  slots are used: in one time slot
both users are active  while in the other time slot only one of them is transmitting.
As can be seen from this figure, both NOMA and OMA are suboptimal
when compared with the case where NOMA and OMA are combined with.

\subsubsection{Interfering MAC and BC}
Consider a mutually interfering two-cell network in the uplink, where each cell includes one MAC.
Assume that  only  one of the
transmitters of each MAC (typically the closest one to the cell-edge) is interfering with the BS of the other MAC.
In this  network,  the  interfering transmitters can employ HK coding, similar to that used in the IC,
while the non-interfering transmitters in each MAC employ
single-user  coding. This  NOMA-based transmission results in an inner bound which
is within a one-bit gap of the capacity region \cite{pang2016approximate}.
Likewise, one can use interfering BC   to model
a mutually interfering two-cell downlink network.

Despite years of intensive research,  finding optimal uplink and downlink transmit/receive strategies for multi-cell networks remains rather elusive. 
In fact, as discussed earlier, even
for a much simpler case of the two-user IC, the optimal
coding strategy is still unknown.
Nonetheless,  fundamental results from information theory as a whole suggest that NOMA-based techniques
result in a superior rate region when compared with OMA.

It should be highlighted that, despite the above insight from information theory,
OMA techniques have been used in the cellular networks from 1G to 4G, mainly to avoid interference due to its simplicity.\footnote{{For 3G, wideband code-division multiple access (WCDMA) was adopted, wherein \textit{orthogonal} channelization codes  are used within a cell, yet \textit{quasi-orthogona}l scrambling  codes are  used to reduce the inter-cell interference.}}
In addition, the lack of understanding of optimal strategies for multi-cell networks has motivated  pragmatic approaches in which  interference is simply treated as noise.

\section{Single-Cell NOMA: A Review}
\label{sec:SNOMA}

As explained in Section~\ref{sec:theory}, the basic theory of NOMA has been around for several decades. However,
a new wave of research on NOMA has been motivated by the advance of processors which make it possible to implement SIC at the user equipment.
Saito et al. \cite{saito2013non} first observed the potential of NOMA for 5G
systems. They showed that NOMA can improve system throughput and  user-fairness over orthogonal frequency division multiple access (OFDMA).
Since then, NOMA has attracted considerable attention from both  industry and academia.
To make this concept more practical, several issues like user-pairing,
power allocation, and SIC implementation issues have been studied in \cite{ding2016impact}.
NOMA has also been  considered in the 3GPP LTE-A
systems under the name of multi-user superposition transmission \cite{NOMA3GPP}.

The performance of NOMA can be further boosted in multi-antenna networks.
MIMO-NOMA solutions exploit multiplexing and diversity gains
to improve outage probability and throughput, by converting the MIMO
channel into multiple parallel channels \cite{ding2015application}.

\section{Multi-Cell NOMA Solutions}
\label{sec:MNOMA}

\begin{figure*}[t]
\centerline
{\centering
\includegraphics[height=7.7cm] {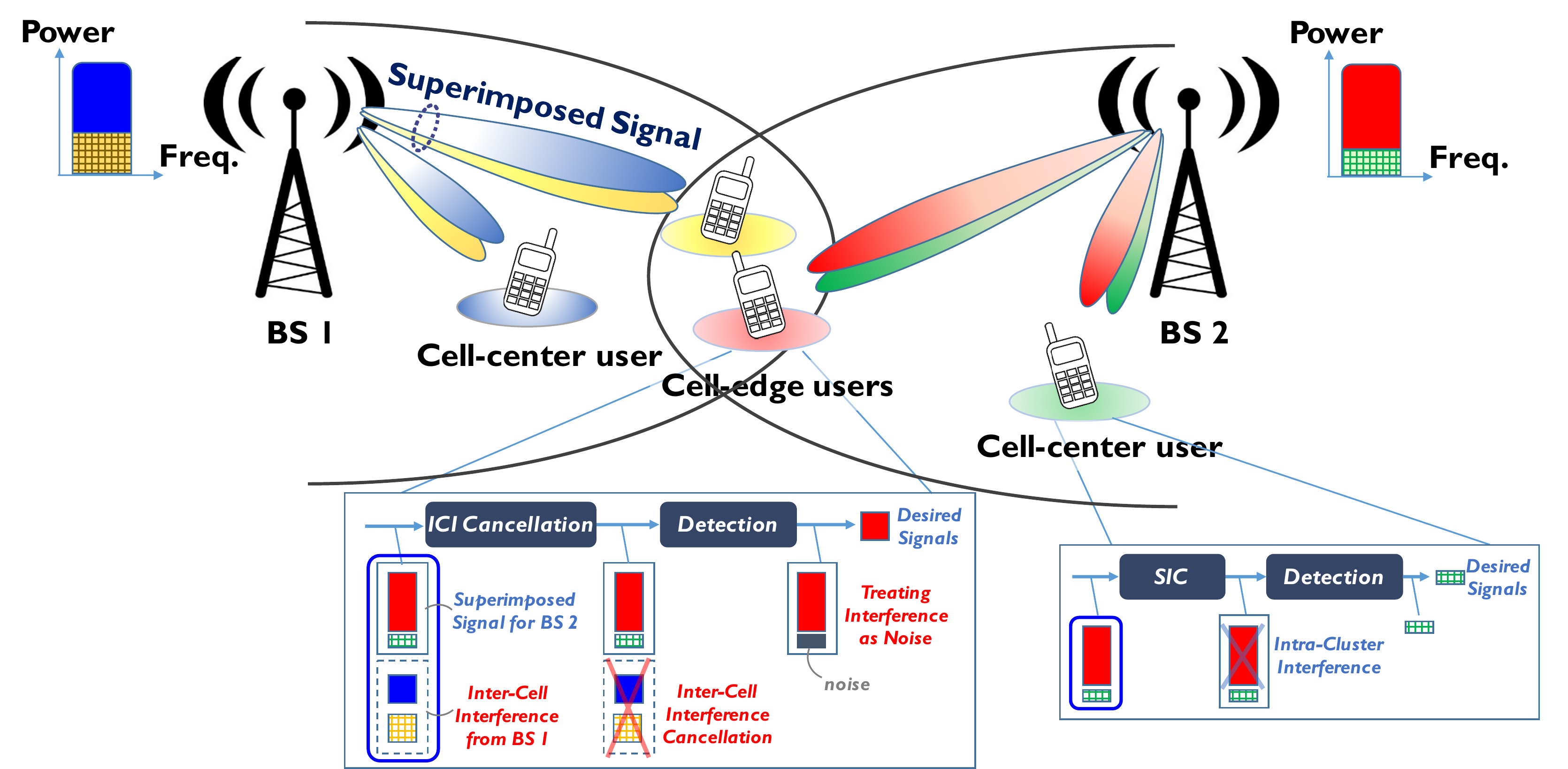}}%
\caption{An illustration of multi-cell NOMA networks.}
\label{fig2}
\end{figure*}

In this section, we discuss recent research that combines interference management
approaches with NOMA, called multi-cell NOMA.
As illustrated in Fig.~\ref{fig2}, ICI is the main
issue in multi-cell NOMA networks, as it reduces a cell-edge user's performance.
This is in contrast with single-cell NOMA, which aims at improving the user-fairness.
Multi-cell techniques are used to harness the effect of ICI.

Multi-cell techniques can be categorized into coordinated scheduling/beamforming (CS/CB)
and joint processing (JP) \cite{lee2012coordinated}. This classification is based on whether the data
messages desired at the users should be shared among multiple BSs or not. These techniques can be combined with NOMA.
For NOMA-CS/CB, data for a user is only available at and transmitted from a single BS.
In contrast, NOMA-JP relies on data sharing among more than one BS.



\subsection{NOMA with Joint Processing}
In NOMA-JP, the users' data symbols are available  at more than one BS. Based on the number of active BSs that serve a user, we can further divide NOMA-JP into two classes: NOMA-joint transmission (JT) and NOMA-dynamic cell selection (DCS).

\subsubsection{NOMA-JT}
\begin{figure*}[t]
	\centerline
	{\centering
		\includegraphics[height=14.0cm] {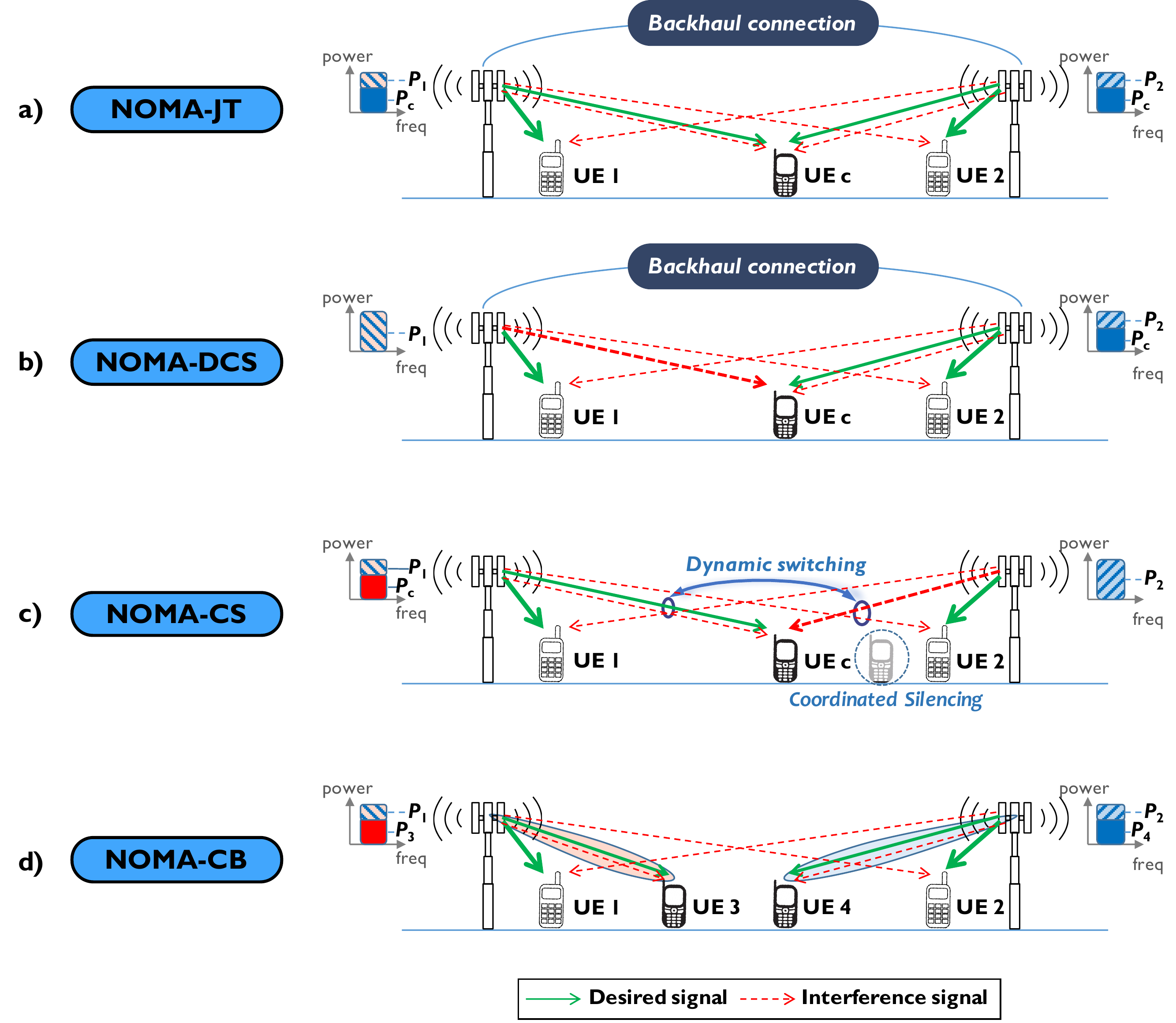}}%
\caption{Multi-cell NOMA solutions: a) NOMA-JT, b) NOMA-DCS, c) NOMA-CS, d) NOMA-CB }
	\label{fig3}
\end{figure*}

This approach requires multiple BSs to simultaneously serve a user using a shared wireless resource instead of acting as interference to each other. This significantly improves the quality of the received signal at cell-edge users at the cost of slightly diminished rates for cell-center users.
This cooperative setting is  similar to a single-cell NOMA as the ICI for cell-edge users can be completely canceled using network MIMO techniques \cite{han2014nnoma}.
Such an approach usually relies on global channel state information (CSI) at all transmitters, which results in excessive backhaul overhead. To overcome the CSI sharing overhead for NOMA-JT, a coordinated superposition coding (CSC) scheme for a two-cell downlink network  was introduced in \cite{choi2014non}. In this scheme, each cell-center user is served by its corresponding BS while the cell-edge user is served by both BSs, as shown in  Fig.~\ref{fig3}. Specifically, two BSs transmit Alamouti coded signals to a cell-edge user to achieve a higher transmission rate, while each BS also transmits signals to the cell-center user.
It has been shown that the coordination between two cells allows NOMA to provide a common cell-edge user with a reasonable transmission rate without  sacrificing cell-center users' rates.
Let $P$ and $P_\mathsf{c}$ be the powers of the cell-center and cell-edge users' messages per cell, respectively. {Assume that $\gamma_{i,m}=|h_{i,m}|^2$ for $i\in\{1,2,\mathsf{c}\}$ and $m\in\{1,2\}$, where $h_{j,m}$ and $h_{\mathsf{c},m}$ denote the channel coefficients to the cell-center user in cell $j$ and the common cell-edge user from BS $m$, $\forall j,m\in\{1,2\}$, respectively. The sum-rate of NOMA-JT given by  $R_1\! +\! R_2 \!+\! R_{\mathsf{c}}$ (sum of rates for cell-center users $R_1$ and $R_2$, and a common cell-edge user $R_\mathsf{c}$) where $ R_1 = \mathcal{C}(\frac{\gamma_{1,1}P}{{\gamma}_{1,2}P+1})$, $R_2 = \mathcal{C}(\frac{\gamma_{2,2}P}{{\gamma}_{2,1}P+1})$, and $R_{\mathsf{c}} = \min\big\{\mathcal{C}(\frac{({\gamma}_{1,1}+{\gamma}_{1,2}){P_\mathsf{c}}}{({\gamma}_{1,1}+{\gamma}_{1,2})P+1}),\mathcal{C}(\frac{({\gamma}_{2,1}+{\gamma}_{2,2}){P_\mathsf{c}}}{({\gamma}_{2,1}+{\gamma}_{2,2})P+1}),\mathcal{C}(\frac{({\gamma}_{{\mathsf{c}},1}+{\gamma}_{{\mathsf{c}},2}){P_\mathsf{c}}}{({\gamma}_{{\mathsf{c}},1}+{\gamma}_{{\mathsf{c}},2})P+1})\big\}$.} Note that the last term comes from the condition that cell-edge user's message has to be decoded by that user and also cell-center users in both cells in order to operate SIC.

\subsubsection{NOMA-DCS} In this case, the user's data is shared among multiple BSs, but it is transmitted only from one  selected  BS. Note that the transmitting BS can be dynamically changed over time  by using  order statistics. Suppose $|h_{c,2}|^2 > |h_{c,1}|^2$; then, BS 2 becomes  the sole serving BS for a cell-edge user until the order statistics is not changed. That is, only BS 2 employs NOMA strategy to support a pair of cell-edge and cell-center users at the same time while BS 1 serves only its corresponding cell-center user (see Fig. 3). Since BS~1 employs OMA instead of NOMA, rate expressions for NOMA-JT, except for $R_2$, should be modified for NOMA-DCS as ${ R_1=\mathcal{C}(\frac{\gamma_1P}{\bar\gamma_1(P+P_\mathsf{c})+1})}$ and $ { R_{\mathsf{c}}=\min\{ \mathcal{C}(\frac{\gamma_2P_\mathsf{c}}{(\gamma_2+\bar{\gamma_2})P+1}),\mathcal{C}( \frac{\gamma^{c}_2P_\mathsf{c}}{(\gamma^{c}_1+\bar{\gamma^{c}_2})P+1})\}}$ since user 1 does not use SIC for NOMA transmission.

\subsection{NOMA with Coordinated Scheduling/Beamforming}
The designs of CS/CB for NOMA differ from those of JP in that the users' data  are not shared among the BSs. However, the cooperating BSs still need to exchange global CSI and cooperative scheduling information via a standardized interface named X2. This may result in a non-negligible overhead especially for high mobility cell-edge users. This subsection briefly discusses how to apply CS or CB to NOMA to tackle ICI problem. The illustration of NOMA-CS/CB is shown in Fig. 3.
\subsubsection{NOMA-CB}
In this case, data for a user is only available at one serving BS, and  the beamforming decision is made with coordination that relies  on global CSI. The authors in \cite{shin2016coordinated}  proposed two novel \textit{interference alignment} (IA)-based CB methods in which two BSs jointly
optimize their beamforming vectors in order to
improve the data rates of cell-edge users by removing ICI. Both algorithms aim to choose the transmit/receive beamforming vectors to satisfy the zero ICI as well as zero inter-cluster interference.
These algorithms are termed interfering channel alignment (ICA)-based CB and IA-based CB. The former requires global CSI  at the BS. However, the latter only requires the knowledge of cell-edge users' serving channel at the BS but with a slightly large number of antennas to compensate for the lack of interfering channels' knowledge. In particular, when the number of users is sufficiently large, it turns out that the number of extra antennas required  for the latter scheme becomes negligible. Moreover, the transmit and receive beamforming vectors for uplink multi-cell NOMA also can be directly obtained by  uplink-downlink duality.
\subsubsection{NOMA-CS}
The key idea of NOMA-CS is to allow geographically separated BSs to coordinate scheduling to serve  NOMA users with less ICI so as to ensure the proper QoS  of cell-edge users. To guarantee the required data rate of the cell-edge users in a cell, the adjacent BS may decide not to transmit a superimposed message to a set of NOMA users, but just a dedicated message to a single cell-center user as in Fig.~\ref{fig3}. However, such a NOMA-CS scheme is formulated as a combinatorial optimization, which is  NP-hard. Therefore, a simple scheduling algorithm is indispensable in order to determine a set of NOMA users scheduled in each BS within a certain scheduling interval.

\begin{table*}[b]\begin{center}
		\caption{A comparison of different multi-cell NOMA solutions.}
		\begin{tabular}{l | a | b | a | b| a }
			\hline
			\rowcolor{LightCyan}
			\mc{1}{}  & \mc{1}{NOMA-CS} & \mc{1}{NOMA-CB}  & \mc{1}{NOMA-DCS} & \mc{1}{NOMA-JT}  \\
			\hline
			\# of transmission  points & 1 & 1 & 1 (Dynamic) & $\geq 2$ \\		 \hline
			Shared information& CSI, Scheduling& CSI, BF& CSI, Data & (CSI), Data, BF \\ \hline
			\centering Backhaul type & Non-ideal  & Non-ideal  & Ideal  & Ideal  \\ \hline
			Total \# of supported users& $\ll 4K$& $ 4(K-1)$ &  $3K$ & $ 3K\,$ (or $4K$) \\ \hline
			References & &\cite{shin2016coordinated}&\cite{choi2014non}&\cite{han2014nnoma}, \cite{choi2014non}\\\hline
		\end{tabular} \end{center}
	\end{table*}

A summary of different multi-cell NOMA techniques is provided in Table 1. In addition, we compare the number of supported users by different NOMA schemes according to the number of clusters in each cell and the number of BS/user antennas.  For comparison, we consider  two-cell scenarios and assume that each BS and user has $K$ antennas. Each cell consists of $K$  clusters each having two users. {It should be highlighted that single-cell NOMA can support $2K$ users}\cite{ding2015application}. {In contrast, single-cell OMA can serve only $K$ users since the number of served users is  limited by the number of antennas at the BS}\cite{tse2005fundamentals}.


\section{Practical Challenges for Multi-Cell NOMA}
\label{sec:challenges}
\subsection{SIC Implementation Issues}
As seen in Section~\ref{sec:theory}, SIC is at the heart of NOMA, and NOMA achieves the capacity region of the downlink and uplink channels (in a single-cell network)
and  the best rate region in the multi-cell setting. SIC, however,  suffers from  several
practical issues, such as:

\subsubsection{Hardware Complexity}
SIC implies that each user has to decode information intended for all
other users before its own  in the SIC decoding order.
This causes the complexity of decoding to scale with the number of users in the cell.
To reduce the complexity, we can divide users in to multiple clusters and apply
encoding/decoding within each cluster. Then, the complexity would be reasonable enough to be
handled thanks to the advance of processor technologies during past decades.  In fact, 3GPP LTE-A recently includes a new category of relatively complex user terminals, named network assisted interference cancellation and suppression (NAICS).

\subsubsection{Error propagation}
Error propagation means that if an error occurs in decoding  a certain user's signal, all other users
after this user in the SIC decoding order will be affected and their signals are likely to be decoded incorrectly. The side effect can  be compensated by using stronger codes provided that the number of users is not very large.{\footnote{{By making use of  implementable near-capacity achieving AWGN channel codes (such as LDPC codes), we can get closer to the capacity region in practice.}}}

\subsection{Imperfect CSI}
Without perfect CSI at the user side it is not possible to completely remove the effect of the other users' signals from the received signal, which results in error propagation. Moreover, without perfect CSI about the interfering links at the BS, a joint precoder that guarantees no ICI is not known yet. In this regard,
new beamforming designs which are robust to CSI errors must be developed for multi-cell NOMA.

\subsection{Multi-User Power Allocation and Clustering }
Power allocation and clustering are  important factors that determine the performance gain of NOMA.
To explain the effect of these factors, consider the simple case of single-cell two-user NOMA and assume
  that $\beta P$ and $(1-\beta) P$ are the powers allocated to user~1 and
user~2. As described in Section~\ref{sec:theory}, by varying $\beta$ different points on the NOMA curve can be achieved.
Therefore, $\beta$ determines the rates for the users. This implies that with
power allocation we can manage the system throughput and user-fairness. {If there are more than two users
in one cell, from the theory we know that all users' signals  should be superimposed together;
i.e., having one cluster maximizes the system throughput} . However, in practice, having only one cluster can result in a serious performance degradation due to SIC error when there are many users in each cell.
A suboptimal, but more practical, solution is to have multiple clusters per cell.
However, it is still very hard to find the optimal clustering for a given number of clusters and the optimal solution is unknown.
In multi-cell networks, ICI comes in  which makes clustering, and  power allocation
even harder. It should be noted that in the  multi-cell case, even when no SIC error assumed,
the optimal clustering and power allocation solutions are not known. Therefore, clustering and power allocation algorithms with reasonable complexity and good performance are inevitable to implement NOMA in practical cellular systems.

\subsection{Operation with FFR}

The basic idea of {fractional frequency reuse} (FFR)  
 is to split a cell's bandwidth into multiple subbands
and \textit{orthogonally}  allocate  subbands for the cell-edge regions of the adjacent cells.
This concept is in contrast with  NOMA wherein orthogonalization is avoided due to its suboptimality.
Despite being theoretically suboptimal, FFR is important as it offers a simple approach for ICI management without requiring CSI. Thus, it is important
to investigate methods that can  bring NOMA and  FFR-based networks together.
A simple idea to make use of both FFR and NOMA  is to apply NOMA in the cell-center band and cell-edge band separately, which would pair
 cell-center users together (in the cell-center band) and  cell-edge users together (in the cell-edge band).  However,
such users are not expected to have very different channel conditions, and any NOMA gain may not be noteworthy.
Another idea is to pair a user from the  cell-center region with a user from the cell-edge region in the cell-edge band to avoid ICI.
Such a pairing will reduce cell-edge users' rates as their specific bands can be shared by the cell-center users too, which sacrifices the cell-edge users' rates.
This, in turn,  deteriorates user-fairness.

\subsection{Security}
The fact that in a NOMA-based transmission the user with better channel condition is able to decode the
other user's signal brings new security  concerns.
Upper-layer security approaches (e.g., cryptographic) are still relevant  since only the legitimate user  has a key to  decode its message.
Nonetheless, physical layer security schemes are of interest but cannot be easily applied to the new environment.

	\section{Performance of Multi-Cell NOMA}
	\label{sec:performance}

In order to observe the potential gain of  NOMA, numerical analysis is performed under a realistic multi-cell environment. In our simulation, we consider a two-cell downlink cellular network.
As a performance metric, we use the cumulative distribution function (CDF) of the user throughput, and the individual user throughputs for the cell-center and cell-edge users.
Detailed simulation parameters are provided in Table II. In our simulation, following schemes are considered: OMA, OMA-FFR, NOMA, NOMA-TDM, NOMA-JT, and NOMA-CB.
In OMA-FFR, FFR 
is used in controlling ICI on top of OMA transmission. Due to the effect of FFR, the cell-edge users experience no ICI while the cell-center users receive interference from other cell. In OMA, single-cell operation \cite{ding2015application} is applied by treating all ICI as noise.
Compared to OMA-FFR, ICI is a significant issue especially for cell-edge users, resulting in a severe SNR loss. Since it is inherently difficult to apply FFR in NOMA-based schemes as discussed in Section~\ref{sec:challenges}, NOMA-TDM and NOMA schemes are considered. NOMA-TDM refers to a NOMA scheme that allows users in different cells to share one resource block via some form of orthogonalization, but  NOMA simply acts as a single-cell operation \cite{ding2015application} by treating the ICI as noise. In NOMA-CB and NOMA-JT, two BSs jointly optimize their beamforming vectors in order to mitigate ICI \cite{shin2016coordinated, choi2014non}.


\begin{figure*}
	\centering
	\hspace{-12mm}
	\begin{subfigure}[Spectral efficiency as a function of cell-edge user's location]
		{\centering
			\includegraphics[height=7.5cm] {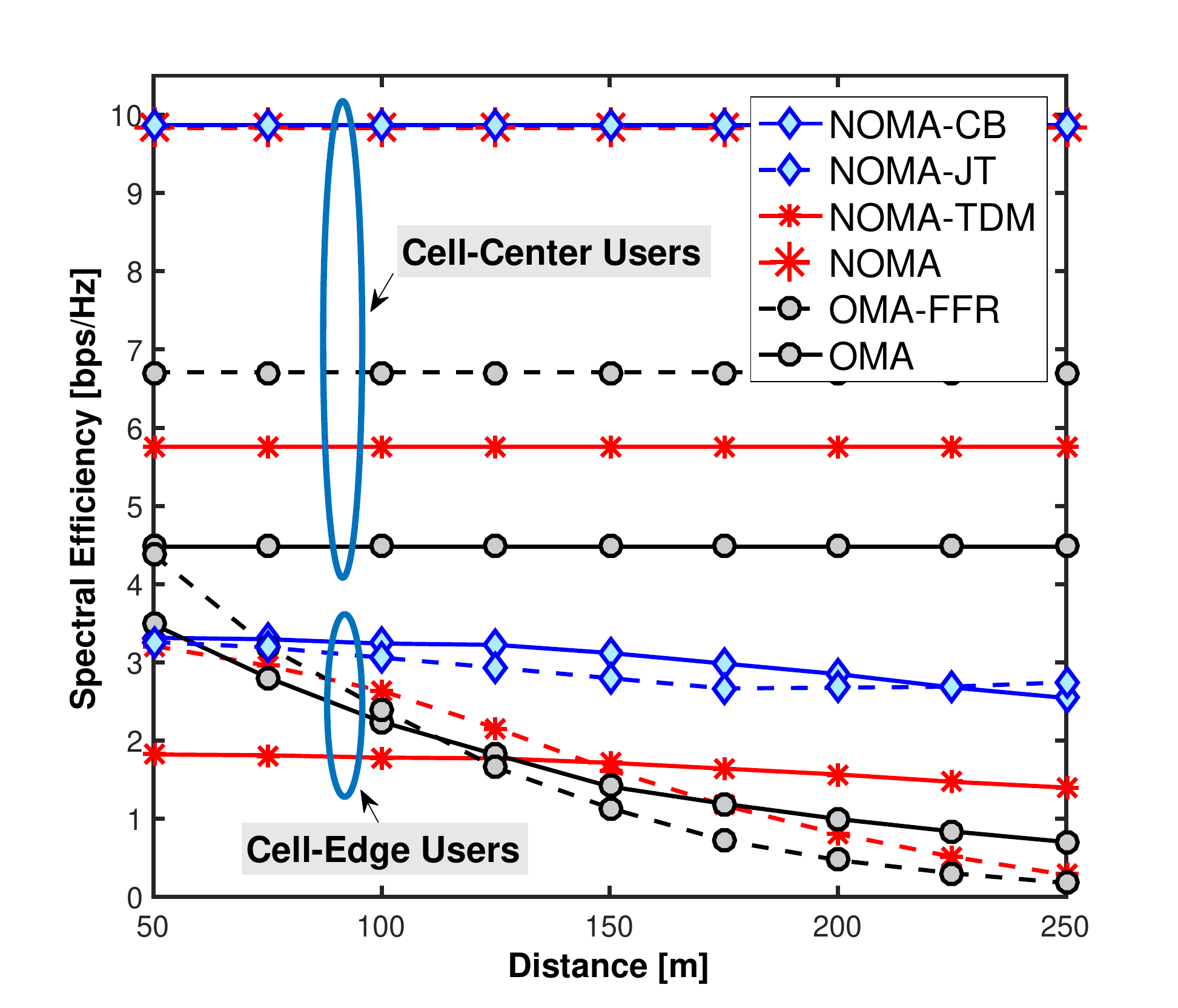}
			\label{fig4:subfig1} }
	\end{subfigure}
	\hspace{-6mm}
	\begin{subfigure}[Individual data rate CDF for random user deployments]
		{\centering
			\includegraphics[height=7.5cm] {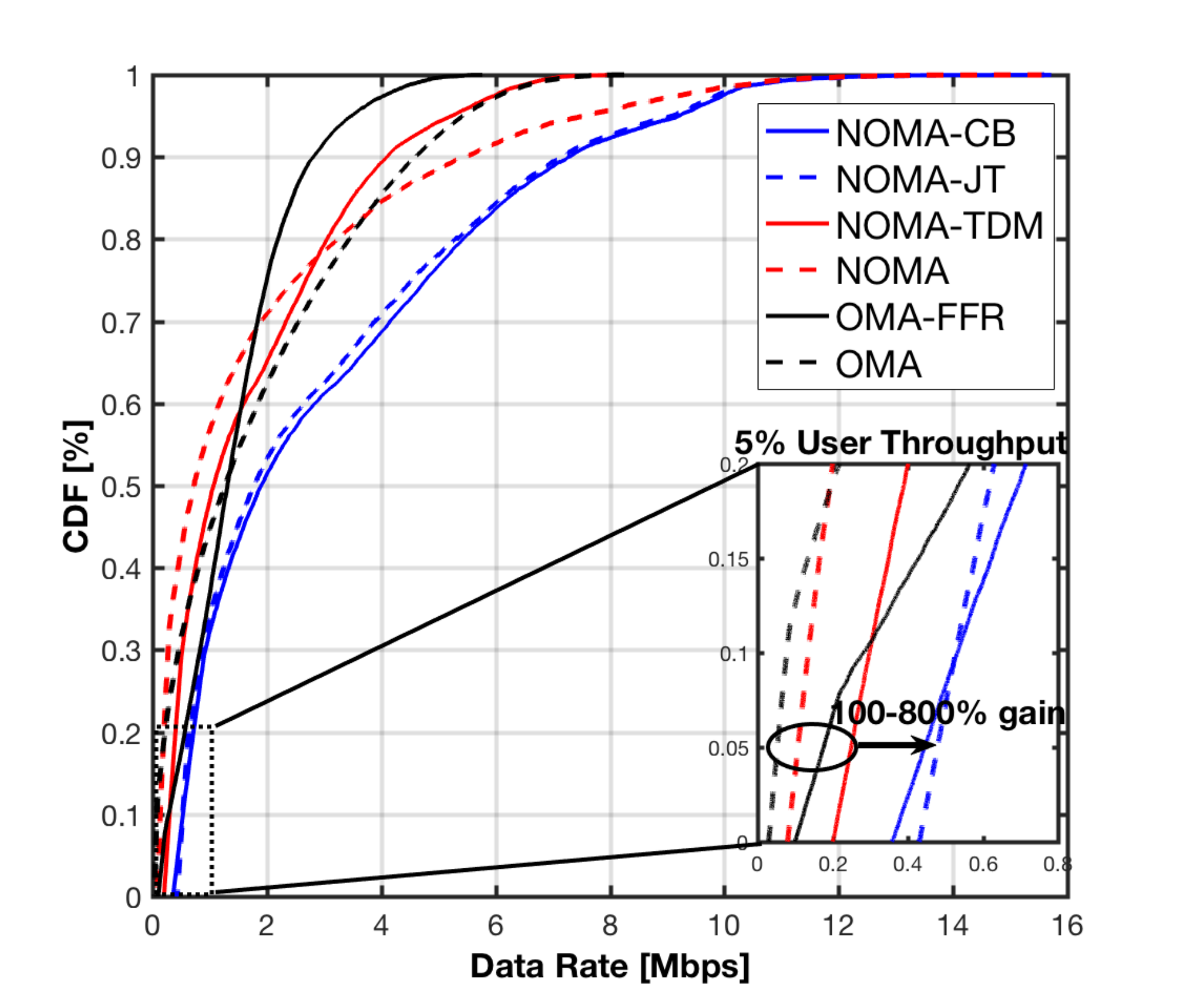}
			\label{fig4:subfig2} }%
	\end{subfigure}
	\hspace{-7mm} \caption{Performance comparison  of different transmission schemes in multi-cell downlink networks}
	\label{fig4:model}
\end{figure*}

In Fig. 4(a), we plot the performance of the cell-center and cell-edge users. Generally, the performance of OMA and NOMA decreases significantly with the location of cell-edge users since ICI mitigation is not considered. On the other hand, NOMA-TDM and OMA-FFR divide resources to support multi-cell environment, thus the rates of cell-edge users are improved compared to the single-cell operation schemes, such as OMA and NOMA.  NOMA-CB can fully exploit all the resources to support all the users, and shows a twice increased performance compared to NOMA-TDM. NOMA-JT  shows a performance almost similar to that of NOMA-CB, but its gain increases as the cell-edge user gets closer to the boarder of the cell. This is because the cell-edge user can take advantage of the link from the neighboring BS to improve its SNR via data sharing.
Note that the cell-edge user performance of OMA-FFR and OMA is even better than that of NOMA-CB when the location of cell-edge user is relatively close to the BS, due to the inherently  remaining inter-user interference of cell-edge user from NOMA transmission \cite{tabassum2016non}. This phenomenon accounts for a  motivation behind  user-pairing for NOMA to be implemented in practice.

In Fig.~\ref{fig4:subfig2}, we plot the CDF of the user throughput. It can be seen that NOMA-CB and NOMA-JT achieve the best performance for any throughput (e.g., 5$\%$-tile CDF point and average for user throughput). This is because ICI is effectively controlled by exploiting the multi-cell NOMA transmissions.
 As a matter of fact, OMA and NOMA are advantageous to the cell-center users' throughput due to full usage of a resource block while these are limited by severe ICI especially for the cell-edge users. {On the contrary}, {NOMA-TDM} and OMA-FFR are deployed to overcome ICI by further splitting a  resource block into two parts (i.e., two cells) with sacrificing cell-center users' throughput.

\begin{table*}
	\begin{center}
		\caption{Simulation Parameters}
		\label{tab:simulation}
		\begin{tabular}{l|l} \hline
			Cell layout & 2 Cells \\ \hline
			Cell radius & 0.25 Km \\ \hline
			Path loss exponent & 4 \\ \hline
			Channel model & Rayleigh fading model \\ \hline
			Channel estimation & Ideal \\ \hline
			Number of transmitter antennas & 4 \\ \hline
			Number of receiver antennas & 4 \\ \hline
			Number of clusters per cell & 4 \\ \hline
			Number of users per cluster & 2 \\ \hline
			Users' locations & Randomly generated and uniformly distributed within the cell  \\ \hline
			User pairing & cell-center user from the disc with radius 0.125 Km \\
			& cell-edge user from the ring \\ \hline
			Transmission power & 10 W\\ \hline
			Noise power spectral density & $10^{-10}$ W/Hz \\ \hline			
			Maximum number of multiplexed UEs & 1 (OMA), 2 (NOMA) \\ \hline
		\end{tabular}
	\end{center}
\end{table*}


\section{Conclusion} \label{sec:Conclusions}
%

In this article, we described the theory behind NOMA in single-cell (both uplink and  downlink) and multi-cell networks.
This is followed by an up-to-date literature review of interference management techniques that apply NOMA in multi-cell networks.
Numerical results  show the significance of  interference cancellation in NOMA.
We have also highlighted  major practical issues and challenges that arise in the implementation of multi-cell NOMA.

\typeout{}
\bibliography{noma-bib}
\bibliographystyle{ieeetr}
\end{document}